\documentclass{ws-procs975x65}
\usepackage{amssymb}

\begin{document}

\title{Emergence of nonlinearity and plausible turbulence in accretion disks via hydromagnetic 
transient growth faster than magnetorotational instability}
\author{Sujit K. Nath$^*$ and Banibrata Mukhopadhyay$^{**}$}

\address{Department of Physics, Indian Institute of Science,\\
Bangalore, Karnataka 560012, India\\
$^*$E-mail: sujitkumar@physics.iisc.ernet.in, 
$^{**}$E-mail: bm@physics.iisc.ernet.in\\
www.iisc.ernet.in}

%

\begin{abstract}
We investigate the evolution of hydromagnetic perturbations in
a small section of accretion disks. It is known that molecular viscosity is negligible 
in accretion disks. 
Hence, it has been argued that Magnetorotational Instability (MRI) is responsible 
for transporting 
matter in the presence of weak magnetic field. 
However, there are some shortcomings, which question effectiveness of MRI.
Now the question arises, whether other hydromagnetic effects, e.g. transient growth (TG), can 
play an important role to bring nonlinearity in the system, even at weak magnetic fields. Otherwise, 
whether MRI or TG, which is primarily responsible to reveal nonlinearity to make the flow 
turbulent? Our results prove explicitly that the flows with high Reynolds number ($R_e$), which is the case of 
realistic astrophysical accretion disks, exhibit nonlinearity by best TG of perturbation modes 
faster than that by best modes producing MRI. 
For a fixed wavevector, MRI dominates over transient effects, only at low $R_e$, lower than 
its value expected to be in astrophysical accretion disks, and low magnetic fields. This seriously questions (overall) 
persuasiveness of MRI in astrophysical accretion disks.
\end{abstract}


\bodymatter


\section{Introduction}
Accretion disks are found in 
active galactic nuclei (AGNs), around a compact stellar object in binary systems, around newly formed 
stars etc. \cite{pringle, shakura}. However, the working principle 
of accretion disks still remains enigmatic to us. Due to its inadequacy of 
molecular viscosity, turbulent viscosity has been proposed to explain the transport of matter towards the
central object. This idea is particularly attractive because of its high $R_e$($\gtrsim 10^{14}$) \cite{bmplb}. 
However, the Keplerian disks, which are relevant to many astrophysical applications, 
are remarkably Rayleigh stable. 
Therefore, linear perturbation cannot induce the onset of turbulence and, consequently,
cannot provide enough turbulent viscosity to transport matter inwards.

With the application of Magnetorotational Instability 
(MRI) \cite{velikhov,chandra} to Keplerian disks, Balbus \& Hawley \cite{bh} showed
that initial seed, weak magnetic field can lead to the velocity and magnetic field perturbations
growing exponentially and reveal the onset of turbulence. 
However, for flows having strong magnetic fields, where the magnetic field 
is tightly coupled with the flow, MRI is not expected to work.
Hence, it is very clear that 
the MRI is bounded in a small regime of parameter values when the field is weak.

It has been argued by several works that transient growth (TG) can reveal nonlinearity and 
transition to turbulence at sub-critical $R_e$ \cite{man,amn,chag,yecko,umurhan,avila,klar}. 
Such sub-critical transition to turbulence was invoked to explain colder purely hydrodynamic accretion flows,
e.g. in quiescent cataclysmic variables, in proto-planetary and star-forming disks, 
the outer region of disks in active galactic nuclei. Note that while hotter flows 
are expected to be ionized enough to produce weak 
magnetic fields therein and subsequent MRI, colder flows may remain to be practically neutral in charge
and hence any instability and turbulence therein must be hydrodynamic.
However, in the absence of magnetic effects, the Coriolis force does not allow
any significant TG in accretion disks in three dimensions, independent of $R_e$ 
\cite{man}, while in pure two dimensions TG could be large at large $R_e$.
However, a pure two-dimensional flow is a very idealistic case. Nevertheless, in the presence
of magnetic field, even in three dimensions, TG could be very large (Coriolis
effects could not suppress the growth). Hence, in a real three-dimensional flow, it is very important
to explore magnetic TG.

In the present paper, we explore the relative strengths of MRI and TG in magnetized 
accretion flows, in order to explain the generic origin of nonlinearity and
plausible turbulence therein. By TG we precisely mean the short-time scale growth due to 
shearing perturbation waves, producing a peak followed by a dip. By MRI we mean the exponential
growth by static perturbation waves. While TG may reveal nonlinearity in the system, depending 
on $R_e$, amplitude of initial perturbation and its wavevector and background rotational profile of the flow, question 
is, can its growth rate
be fast enough to compete with that of MRI? On the other hand, is there any limitation of MRI,
apart from the fact that MRI does not work at strong magnetic fields? Note that some limitations of 
MRI were already discussed by previous authors \cite{mahajan,umurhan-prl1,umurhan-prl2,avila,pessah},
which then question the origin of viscosity in accretion disks.

We show below that the three-dimensional TG dominates over the growth due to MRI modes 
at large $R_e$, bringing nonlinearity in the flows. 
By comparing modes corresponding to static (original MRI) and shearing (TG) waves,
the growth estimates from static MRI waves have already been argued to be misleading \cite{man, amn}.
We will show below that in a shorter time-scale, TG 
reveals nonlinearity into the system. 

We furthurmore explicitly calculate the magnetic field strength above which MRI not working.
We notice that above a threshold 
$R_e$, only TG is sufficient to make the system nonlinear at low magnetic field and there 
is no growth at high magnetic fields. 
The working regime of 
MRI is rather much narrower than it is generally believed.
As TG was argued to be plausible source of nonlinearity in cold disks and 
the growth due to MRI is subdominant compared to TG at high $R_e$ in hot disks, 
TG could be argued to be the source of nonlinearity and plausible turbulence and 
subsequent viscosity, in any accretion disk.


\section{Governing Equations Describing Perturbed Magnetized Rotating Shear Flows}

Within a local shearing box, in Lagrangian coordinate, the perturbed and linearized 
Navier-Stokes, continuity, magnetic induction 
equations and solenoidal condition (for magnetic field) can be written as
\begin{eqnarray}
\dot{\bf {\delta v}}= -\frac{1}{\rho}c_s^2\nabla\delta\rho+\frac{1}{R_e}\nabla^2{\bf \delta v}+2{\bf \delta v \times \Omega}+\frac{1}{4\pi\rho}{\bf B}\cdot\nabla{\bf \delta B}+\Omega {\bf \delta v}\cdot{\mathfrak q},
\label{perturbedns}
\end{eqnarray}
\begin{eqnarray}
\dot{\delta\rho}=-\rho\nabla\cdot{\bf \delta v},
\label{perturbedcont}
\end{eqnarray}
\begin{eqnarray}
\dot{\bf {\delta B}}=\nabla\times({\bf v\times\delta B+\delta v\times B})+({\bf v}\cdot\nabla){\bf\delta B},~~~~
\nabla\cdot{\bf \delta B}=0,
\label{perturbedinduction}
\end{eqnarray}
where ${\bf v}$, ${\bf B}$, $\Omega$, $\rho$, $c_s$ and $R_e$ are the 
background velocity, magnetic field vectors, angular velocity, 
density, sound speed and Reynolds number respectively and the 
quantities with $\delta$ such as ${\bf \delta v}$, ${\bf \delta B}$ etc. are 
the respective perturbed quantities. ${\mathfrak q}$ is the tensor related to the background 
shearing velocity depending on the rotation parameter $q$ \cite{amn}. Here we take the 
background shearing velocity as ${\bf v}=(0,-q\Omega x,0)$, 
where $x$ is the $x$-component of the Cartesian position vector of a fluid element inside the shearing box.


We now work with the incompressible approximation, i.e. $\delta \rho \rightarrow 0$ and $c_s^2\rightarrow\infty$, 
assuming $c_s^2\delta \rho$ to be finite and decompose the 
general linear perturbations into a plane wave form as
\begin{eqnarray}
{\bf \delta v}, {\bf \delta B}\propto exp(i{\bf k}^L\cdot{\bf r}^L),
\label{perturbation}
\end{eqnarray}
when
\begin{eqnarray}
{\bf k}=(k_x,k_y,k_z)=({\bf 1}+\Omega t{\mathfrak q})\cdot {\bf k}^L=(k_x^L+q\Omega tk_y^L,k_y^L,k_z^L),
\label{wavenumber}
\end{eqnarray}
where ${\bf k}$ and ${\bf k}^L$ are the wavevectors in the Eulerian and Lagrangian coordinates respectively and 
$t$ is the time. Now solving 
equations (\ref{perturbedns}), (\ref{perturbedcont}) and (\ref{perturbedinduction}) and using  (\ref{perturbation}) we 
calculate energy of the perturbation and linearity given by 
\begin{eqnarray}
{\cal E}\propto \left({\bf\delta v}^2+\frac{{\bf\delta B}^2}{4\pi\rho}\right), 
~~{\rm Linearity}=\left(\frac{|{\bf\delta v}|}{|{\bf v}|}+\frac{|{\bf\delta B}|} {|{\bf B}|}\right)
\label{totalenergyapp}
\end{eqnarray}
respectively, when $|{\bf\delta v}|/|{\bf v}|$, $|{\bf\delta B}|/|{\bf B}|$ 
at time $t=0$ are respective initial perturbation amplitude (IPA). 
For other details, see Ref. \refcite{skbm}.

\section{Total Energy Growth and Nonlinearity of Perturbations for Different Parameter Values}

The best possible mode for MRI giving rise to the nonlinearity in the system corresponds
to the condition $k_z v_{Az}/\Omega=1$, when $v_{Az}^2=B_z^2/4\pi\rho$, is the Alfv\'en velocity \cite{bh}. The growth
rate for this fastest exponentially growing mode is $3\Omega/4=3/4q$ (since in dimensionless unit $\Omega=1/q$) \cite{bh,man,bhr}. 
Note that an approximate emergence of nonlinearity is defined through the measurement of the quantity 
\textquotedblleft Linearity\textquotedblright as defined in eq. (\ref{totalenergyapp}). 
When Linearity $=1$, the system will start becoming nonlinear which will plausibly lead to turbulence.
For a Keplerian disk ($q=3/2$), the best MRI mode brings in the nonlinearity
at the timescales $\sim 14$ and $23$ rotation times respectively
for IPAs $=10^{-3}$ and $10^{-5}$. However Fig. \ref{pert}a 
shows that there are modes which reveal nonlinearity via TG following eqn. (\ref{totalenergyapp}) at around $3$ and 
$13$ rotational times for IPAs $10^{-3}$ and $10^{-5}$ respectively or even less (Fig. \ref{pert}b), 
which shows faster growth rates than MRI. 
In Fig. \ref{pert}c we show the total energy growth of perturbation for different strengths of 
magnetic fields. Thick and long dashed lines correspond to relatively stronger magnetic fields 
for which there is eventually no energy growth and the system remains linear and stable. 
Dotted and dot-dashed lines correspond to weaker 
magnetic fields for which the total energy starts growing and makes the system nonlinear and plausibly unstable. 
Also it is seen that {\it for a given shearing mode}, in case of weak magnetic fields, nonlinearity comes through MRI 
for low $R_e$, and via TG for high $R_e$, which are the cases for astrophysical accretion disks.

\def\figsubcap#1{\par\noindent\centering\footnotesize(#1)}
\begin{figure}[ht]%
\begin{center}
 \parbox{2.1in}{\includegraphics[width=2.01in]{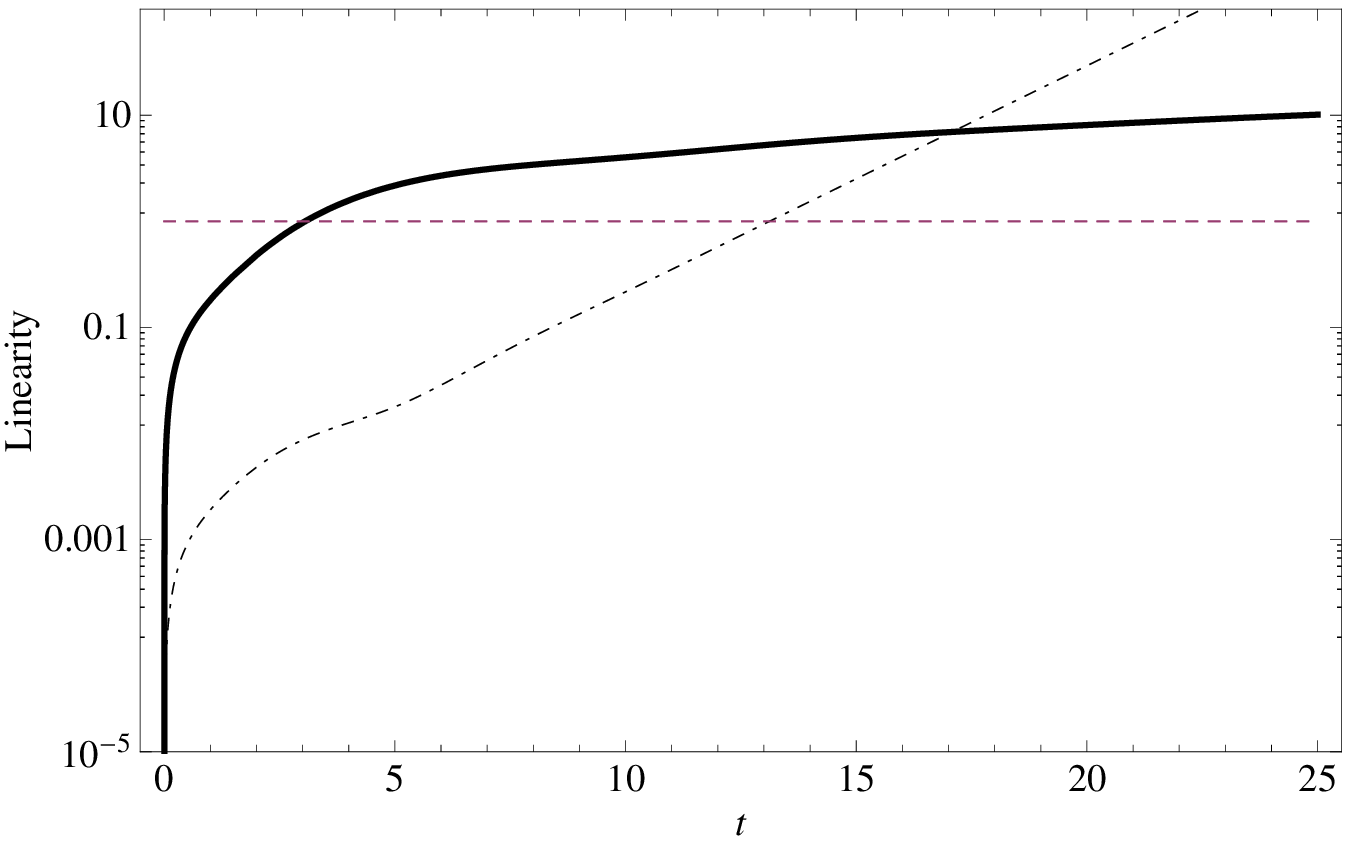}\figsubcap{a}}
 \hspace*{4pt}
 \parbox{2.1in}{\includegraphics[width=2.01in]{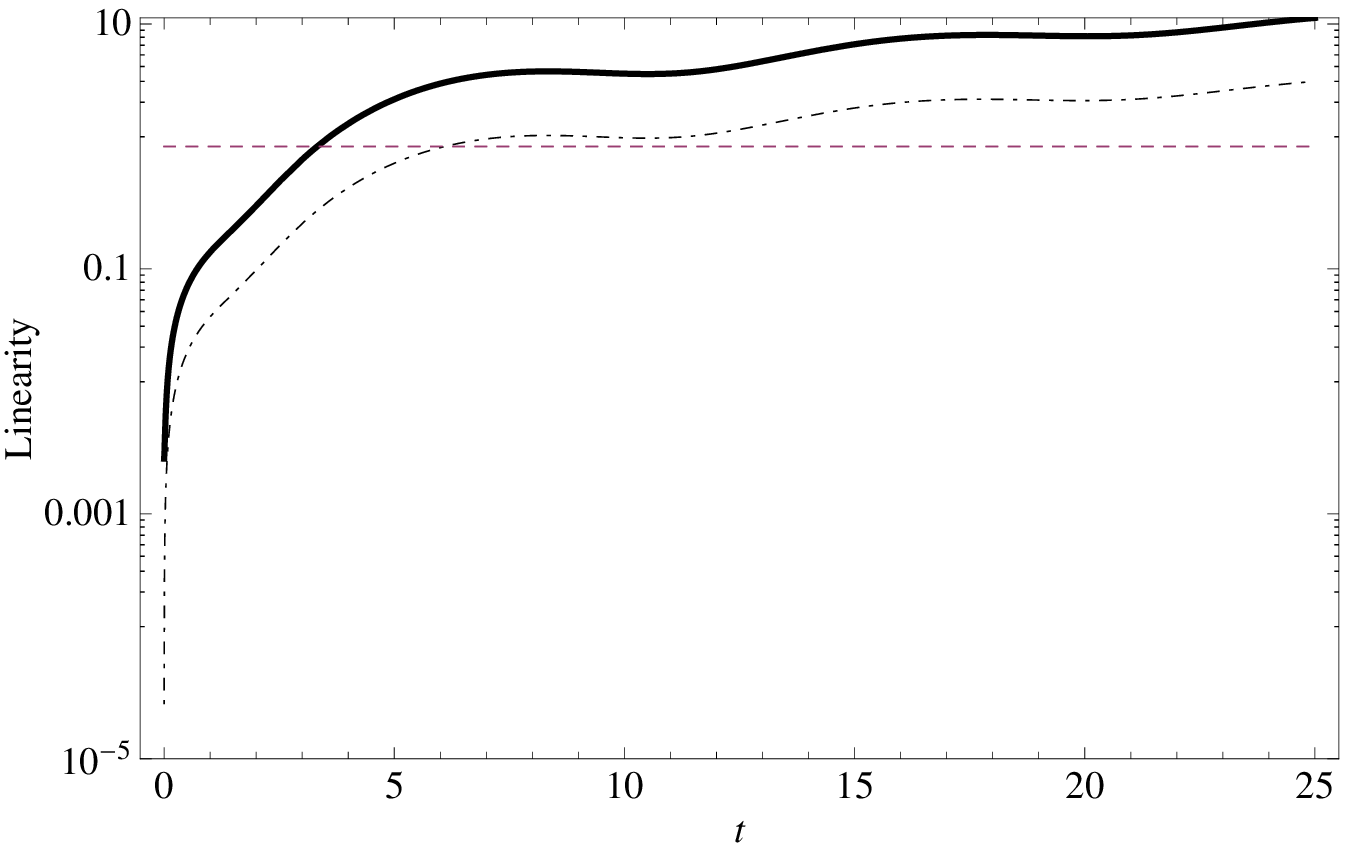}\figsubcap{b}}
 \hspace*{4pt}
 \parbox{2.1in}{\includegraphics[width=1.9in]{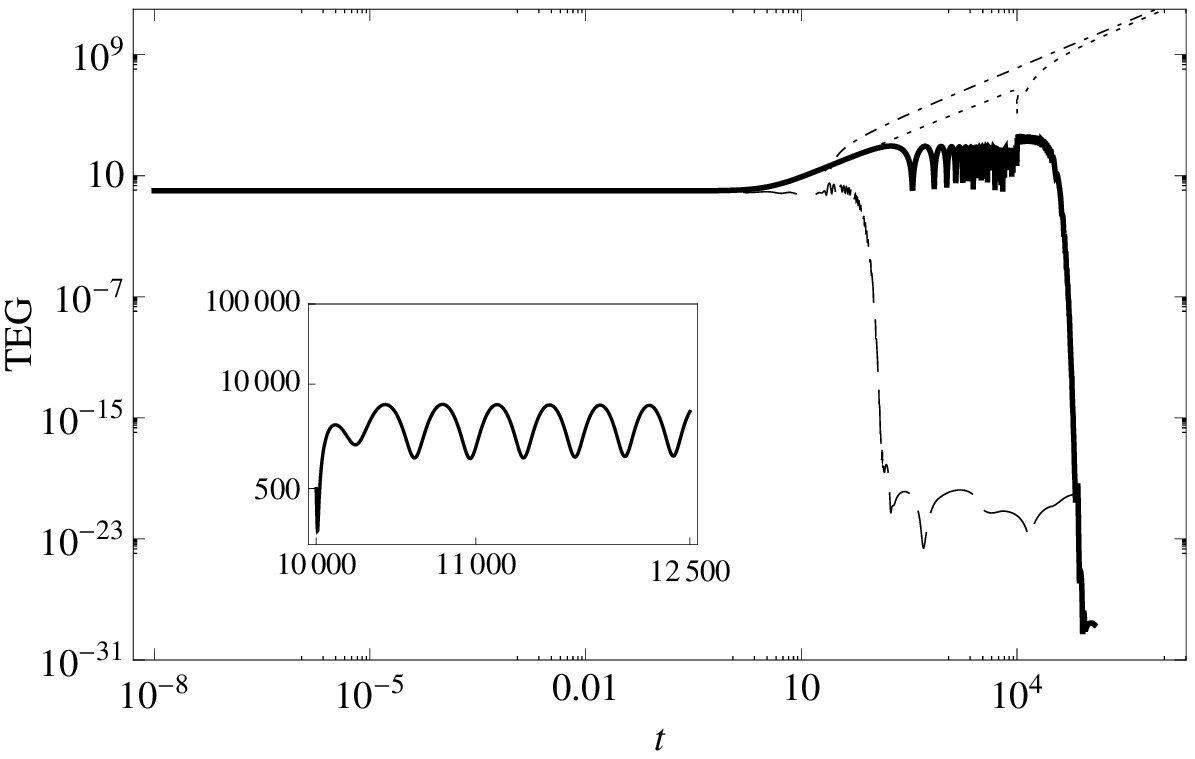}\figsubcap{c}}
 \hspace*{4pt}
 \parbox{2.1in}{\includegraphics[width=2.01in]{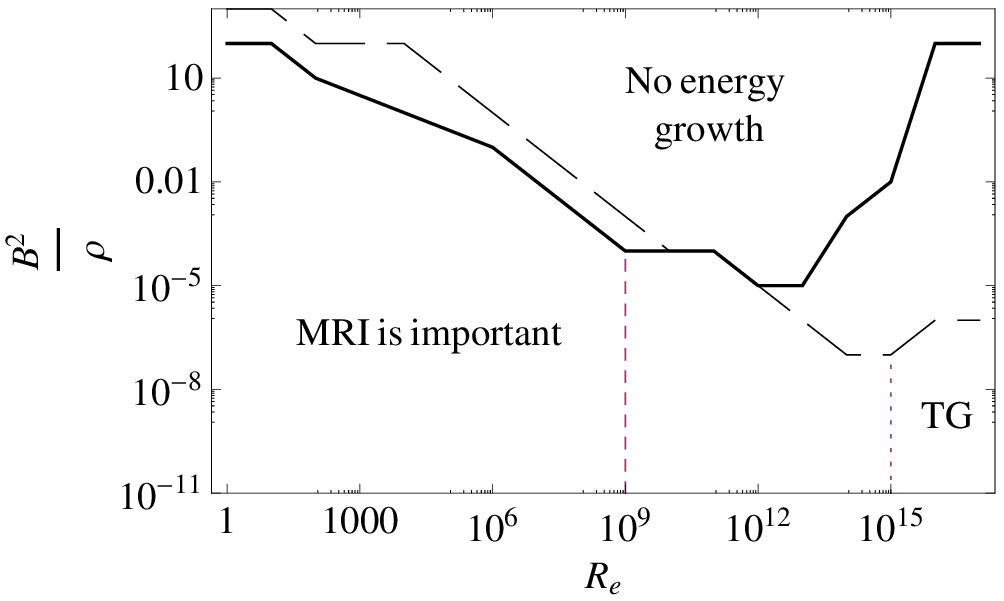}\figsubcap{d}}
 \caption{(a) Nonlinearity via best possible TG and MRI. Thick black line corresponds to
the TG for IPA$=10^{-3}$, $R_e=10^{14}$, $k_{x}^L=-Re^{1/3}$, $k_y=1$, $k_z=90K_x^L$;
dotdashed black line corresponds to the TG for IPA$=10^{-5}$, $R_e=10^{25}$, $k_{x}^L=-Re^{1/3}$, 
$k_y=1$, $k_z=90k_x^L$; red longdashed and dotted lines correspond to the best possible MRI for 
IPA $=10^{-3}$ and $10^{-5}$ respectively. Dashed horizontal line indicates linearity unity. (b) Same as (a), 
but the black thick and dotdashed lines correspond to TG for 
$k_x^L=1$, $k_y=1$, $k_z=100$, $R_e=10^{12}$ 
and $k_x^L=1$, $k_y=1$, $k_z=3000$, $R_e=10^{12}$ respectively. 
(c) Total energy growth for different sets of $R_e$ and ${\bf B}=(0,0,B_3)$ for $k_x^L=-Re^{1/3}$, $k_y=k_z=1$: 
Thick, longdashed, dotted and dotdashed lines correspond
to respectively $R_e=10^{12}~{\rm and}~B^2/\rho=10^{-3}$; $R_e=10^{4}~{\rm and}~B^2/\rho=10$; 
$R_e=10^{12}~{\rm and}~B^2/\rho=10^{-20}$; and $R_e=10^{4}~{\rm and}~B^2/\rho=10^{-20}$. 
Inset confirms that the oscillatory zone of 
thick line is continuous and smooth. (d) Parameter space describing stable and unstable zones, based on the MRI and TG
inactive and active regions, for $k_x^L=-Re^{1/3}$, $k_y=k_z=1$, ${\bf B}=(0,0,B_3)$. Solid and 
longdashed lines are for 
${\rm IPA}=10^{-3}$ and $10^{-5}$ respectively. The dashed and dotted vertical lines at $R_e=10^9$ 
and $10^{15}$ correspond to boundary $R_e$ for the cases
${\rm IPA}=10^{-3}$ and $10^{-5}$ respectively.}
\label{pert}
\end{center}
\end{figure}

\section{Calculation of the Threshold Value of Magnetic Field Strength supporting instability}

Let us estimate the maximum $|\bf B|$ in Gauss supporting nonlinearity, as shown by the solid curve
in Fig. \ref{pert}d. We set the shearing box at $100R_g$ away 
from a $10M_\odot$ black hole. Then we obtain the values of density ($\rho_{100R_g}$) at that location 
to be $\sim 10^{-4}$ gm/cc \cite{shakura}. The background Keplerian 
velocity at that position, for the size of the shearing box, $0.1R_g$, which is consistent with that obtained
for the TG active zone \cite{skbm}, can be obtained as 
$q\Omega L=q\sqrt{GM/R^3}L\sim 10^6$ cm/sec.
We now consider $R_e=10^{12}$ and, hence, 
from the solid line of Fig. \ref{pert}d the corresponding maximum (dimensionless) 
magnetic field supporting nonlinearity is given by $B^2/\rho=10^{-5}$. Therefore,
corresponding actual value of magnetic field is $\sqrt{10^{-5}\rho_{100R_g}(q\Omega L)^2}\sim 30$ Gauss. 
This means, the flow with $R_e=10^{12}$ and 
$|{\bf B}|>30$ 
Gauss, the energy growth of perturbation will decay over time, but for $|{\bf B}|\leqslant 30$ Gauss, TG 
will be sufficient enough to bring nonlinearity in the system, however, still not requiring any growth due to MRI. 
From Fig. \ref{pert}d, 
it is clear that MRI is only important whenever $R_e < 10^9$, whereas for $R_e \geqslant 10^9$, which is the 
favorable zone of $R_e$ for accretion disks, magnetic TG is important than MRI.

\section{Conclusions} 
Here we have shown that, in accretion disks, there are TG modes, which bring nonlinearity faster than the 
best possible MRI mode. We have 
computed the magnetic field strengths for different $R_e$s above which the system will be stable under 
linear perturbation. We have also calculated, for a given shearing mode, an upper bound of $R_e$ 
above which either the system is stable under linear perturbation 
(for high magnetic field strength) or reaches nonlinear regime (for low magnetic field) through magnetic TG (Fig. \ref{pert}d). 
Since astrophysical accretion flows have high $R_e$ ($\gtrsim 10^{14}$) \cite{bmplb}, 
it becomes nonlinear plausibly by magnetic TG.
Hence, MRI is not the sole mechanism to make accretion disk unstable, there is a large area where TG rules,
and explanation of accretion solely via MRI is misleading.\\


\begin{thebibliography}{0}

\bibitem{pringle}
 J.E. Pringle, {\em ARA\&A} \textbf{19,} 137 (1981).

\bibitem{shakura}
 N.I. Shakura and R.A. Sunyaev, {\em Astron. Astrophys.} \textbf{86,} 337 (1973).

\bibitem{bmplb}
B. Mukhopadhyay, {\em Phys. Lett. B} \textbf{721,} 151 (2013).

\bibitem{velikhov}
E. Velikhov, {\em J. Exp. Theor. Phys.} \textbf{36,} 1398 (1959).

\bibitem{chandra}
S. Chandrasekhar, {\em Proc. Nat. Acad. Sci.} \textbf{46,} 53 (1960).

\bibitem{bh}
 S.A. Balbus and J.F. Hawley, {\em Astrophys. J.} \textbf{376,} 214 (1991).

\bibitem{man}
B. Mukhopadhyay, N. Afshordi and R. Narayan, {\em Astrophys. J.} \textbf{629,} 383 (2005).

\bibitem{amn}
N. Afshordi, B. Mukhopadhyay and R. Narayan, {\em Astrophys. J.} \textbf{629,} 373 (2005).

\bibitem{chag}
G.D. Chagelishvili, J.-P. Zahn, A.G. Tevzadze and J.G. Lominadze, {\em Astron. Astrophys.} \textbf{402,} 401 (2003).

\bibitem{yecko}
P.A. Yecko, {\em Astron. Astrophys.} \textbf{425,} 385 (2004).

\bibitem{umurhan}
 O.M. Umurhan and O. Regev, {\em Astron. Astrophys.} \textbf{427,} 855 (2004).

\bibitem{avila}
M. Avila, {\em Phys. Rev. Lett.} \textbf{108,} 124501 (2012).

\bibitem{klar} H.H. Klahr and P. Bodenheimer, {\em Astrophys. J.} \textbf{582,} 869 (2003).

\bibitem{mahajan}
S.M. Mahajan and V. Krishan, {\em Astrophys. J.} \textbf{682,} 602-607 (2008).

\bibitem{umurhan-prl1}
O.M. Umurhan, K. Menou and O. Regev, {\em Phys. Rev. Lett.} \textbf{98,} 034501 (2007).

\bibitem{umurhan-prl2}
E. Liverts, Y. Shtemler, M. Mond, O.M. Umurhan and D.V. Bisikalo, {\em Phys. Rev. Lett.}
\textbf{109,} 224501 (2012).

\bibitem{pessah}
M.E. Pessah and C. Chan, {\em Astrophys. J.} \textbf{751,} 48 (2012).

 \bibitem{bhr}
S.A. Balbus and J.F. Hawley, {\em Rev. Mod. Phys.} \textbf{70,} 1 (1998).
%

 \bibitem{skbm}
S.K. Nath and B. Mukhopadhyay, {\em Phys. Rev. E} \textbf{92,} 023005 (2015).

\end{thebibliography}
\end{document}